\documentclass[conference]{ieeeconf}
\IEEEoverridecommandlockouts
\usepackage{cite}
\usepackage{amsmath,amssymb,amsfonts}
\usepackage{algorithmic}
\usepackage{colortbl}
\usepackage{diagbox}
\usepackage{graphicx}
\usepackage{hyperref}
\usepackage{textcomp}
\usepackage{xcolor}
\usepackage{setspace}
\usepackage{multirow}
\def\BibTeX{{\rm B\kern-.05em{\sc i\kern-.025em b}\kern-.08em
    T\kern-.1667em\lower.7ex\hbox{E}\kern-.125emX}}

\begin{document}
\makeatletter
\newcommand{\linebreakand}{%
  \end{@IEEEauthorhalign}
  \hfill\mbox{}\par
  \mbox{}\hfill\begin{@IEEEauthorhalign}
}
\makeatother

\title{LDPM: Towards undersampled MRI reconstruction with MR-VAE and Latent Diffusion Prior}
\author{Xingjian Tang$^{\dagger}$$^{1,2}$ and Jingwei Guan$^{\dagger*}$$^{1}$ and Linge Li$^{\dagger}$$^{3}$ \\ and Ran Shi$^{4}$ and Youmei Zhang$^{5}$ and Mengye Lyu$^{*1}$ and Li Yan$^{*1}$
\thanks{The authors with $\dagger$ contributed to the work equally and should be regarded as co-first authors.
Corresponding authors are marked with $^*$. Contact E-mail: guanjingwei@sztu.edu.cn.}
\thanks{$^{1}$ Shenzhen Technology University; $^{2}$ Shenzhen University; $^{3}$ Huawei; $    $ $^{4}$ Nanjing University of Science and Technology $^{5}$ Qilu University of Technology
}%
        }
\maketitle
\begin{abstract}
Diffusion models, as powerful generative models, have found a wide range of applications and shown great potential in solving image reconstruction problems.
Some works attempted to solve MRI reconstruction with diffusion models, but these methods operate directly in pixel space, leading to higher computational costs for optimization and inference. 
Latent diffusion models, pre-trained on natural images with rich visual priors, are expected to solve the high computational cost problem in MRI reconstruction by operating in a lower-dimensional latent space.
However, direct application to MRI reconstruction faces three key challenges: (1) absence of explicit control mechanisms for medical fidelity, (2) domain gap between natural images and MR physics, and (3) undefined data consistency in latent space. 

To address these challenges, a novel Latent Diffusion Prior-based undersampled MRI reconstruction (LDPM) method is proposed. 
Our LDPM framework addresses these challenges by: 
(1) a sketch-guided pipeline with a two-step reconstruction strategy, which balances perceptual quality and anatomical fidelity, 
(2) an MRI-optimized VAE (MR-VAE), which achieves an improvement of approximately 3.92 dB in PSNR for undersampled MRI reconstruction compared to that with SD-VAE \cite{sd}, and
(3) Dual-Stage Sampler, a modified version of spaced DDPM sampler, which enforces high-fidelity reconstruction in the latent space.
Experiments on the fastMRI dataset\cite{fastmri} demonstrate the state-of-the-art performance of the proposed method and its robustness across various scenarios.
The effectiveness of each module is also verified through ablation experiments.

\end{abstract}


\section{Introduction}
Magnetic Resonance Imaging (MRI) is a non-invasive medical imaging technique frequently used for disease diagnosis and treatment. 
However, the long scan time limits its broader application.
To this end, k-space undersampling technique is employed to accelerate MRI acquisition. 
High acceleration factors can introduce aliasing artifacts, which need to be removed through reconstruction to achieve diagnostic-quality MRI\cite{fastmri}. Methods like parallel imaging\cite{pi1,pi2,pi3} and compressed sensing\cite{cs1,cs2,cs3,cs4} have been proposed to enhance MRI reconstruction, but they still suffer from limitations like residual artifacts and blurring\cite{joint}.

In recent years, deep learning methods have become mainstream techniques for addressing undersampled MRI reconstruction problems\cite{fastmri,varnet,dudornet,reconformer,swinmr,dagan,jcan}, especially those based on diffusion models (DMs, \cite{ddpm,ncsn,sde}).
Chung et al.\cite{scoremri} proposed an SDE model that showed great reconstruction outcomes in various modalities and diverse body parts. 
Cao et al.\cite{hfs} trained a DM with only high frequency MR k-space to preserve the consistency of the acquired low frequency information. 
Güngör et al.\cite{adadiff} utilized the sensitivity maps to enhance inference performance with large step diffusion. 
Ozturkler et al.\cite{smrd} incorporated automatic hyperparameter selection in the sampling stage of DM to enhance model robustness.
Jiang et al.\cite{ppn} proposed an algorithm to accelerate controllable diffusion models for undersampled MRI reconstruction without requiring paired data or retraining across diverse acquisition parameters.

DM-based methods have demonstrated exceptional performance in reconstructing MR images\cite{roger,tcdiff,nila}. 
Nevertheless, most of these methods operate directly in pixel domains (e.g., image domain and k-space), where optimization and inference are computationally demanding.
Therefore, more lightweight methods are needed to enhance the accessibility of DMs and reduce the significant resource consumption.
The work by Gao et al.\cite{u2mrpd} leverages implicit visual knowledge from a large latent diffusion model pre-trained on natural images and achieves well-generalized MRI reconstruction in an unsupervised way, showing the great potential of latent diffusion models for MRI reconstruction tasks. 

\begin{figure*}
    \centering
    \includegraphics[width=1\linewidth]{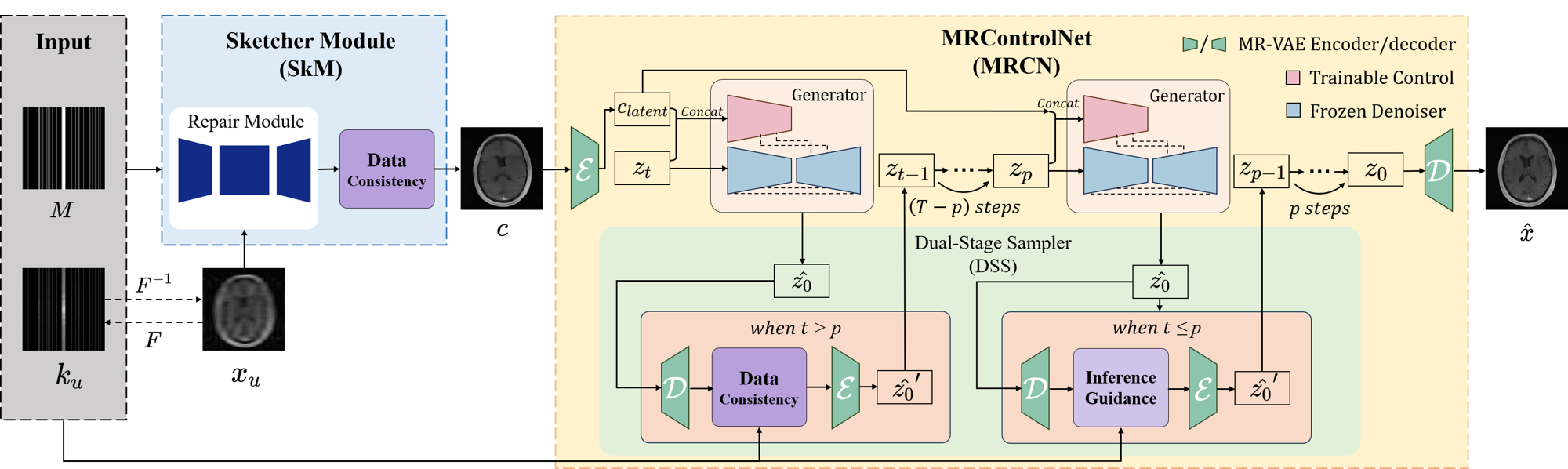}
    \caption{Pipeline of the proposed LDPM method. 1) The Sketcher Module removes the artifacts and generates sketch image $c$ as the conditional input to enhance the fidelity of medical image reconstruction. 2) MRControlNet generates a fully-sampled MRI prediction $\hat{x}$ using MR-VAE and Dual-Stage sampler to ensure both medical fidelity and high visual quality.}
    \label{fig:pipeline}
\end{figure*}

Utilizing latent diffusion models (LDMs)\cite{ldm,sd} is one of the solutions for light-weighted natural image reconstruction\cite{stablesr,refusion,supir,diffuir,pfstorer}. 
However, direct application of LDMs to the task of MRI reconstructions still faces several challenges. 
\textbf{Firstly}, although the vanilla LDM framework is trained on a large-scale dataset and excels in image generation by learning complex data distributions, it may focus more on image visual quality rather than fidelity \cite{stablesr,diffbir} and requires appropriate control.
\textbf{Secondly}, pre-trained variational autoencoders (VAEs)\cite{vae} are utilized in LDMs to map pixel-domain images into the latent space.
As a lossy compression model\cite{lossyae}, the current VAEs pre-trained on natural images may lead to information misinterpretation on MRI images\cite{supir}.
\textbf{Finally}, the widely used operation to ensure the fidelity of MRI reconstruction, data consistency (DC), needs to be modified to adapt to the LDM-based framework and address possible artifacts\cite{u2mrpd}.
In order to fully exploit the advantages of LDM and solve these problems as much as possible, an undersampled MRI reconstruction  method LDPM is proposed.
To conclude, the main contributions of this work are:

\begin{itemize}
\item 

An MR-VAE and Latent Diffusion-Prior based undersampled MRI reconstruction method (LDPM) is proposed.
LDPM is built upon two main modules that function in a cascaded manner, with the first providing control information to ensure the fidelity of medical image reconstruction, and the second module refining details to enhance visual quality. 
The proposed method demonstrates state-of-the-art performance on the fastMRI dataset \cite{fastmri}, with verified robustness across different scenarios, demonstrating its potential for real-world clinical applications.

\item 
MR-VAE, a Variational AutoEncoder (VAE) tailored for MRI tasks, is proposed for MRI-related applications. 
MR-VAE facilitates efficient image transfer to the latent domain, and has been proven highly effective in mitigating compression loss for MR images.
Furthermore, integrating MR-VAE with MRI reconstruction models significantly enhances the quality of the reconstruction results, with an improvement of about 3.92 dB in PSNR for the undersampled MRI reconstruction.

\item 
A variant of the spaced DDPM sampler \cite{iddpm}, named the Dual-Stage Sampler is proposed 
to achieve high-fidelity reconstruction while effectively addressing potential artifacts.
The proposed Dual-Stage Sampler enhances the overall performance and reliability of the MRI reconstruction process within the LDM-based architecture.

\end{itemize}

\section{Method}

The proposed MRI reconstruction method LDPM ($\Phi$) aims at reconstructing the fully-sampled image $\hat{x}$ from the undersampled k-space $k_u$ and its undersampling mask $M$ as \eqref{equ:pipeline}.

\begin{equation}
\hat{x}=\Phi(k_u,M)
\label{equ:pipeline}
\end{equation}
$\Phi$ consists of two main modules, namely the sketcher module and MRControlNet, as shown in \ref{fig:pipeline}.
First, a transformer-based sketcher module (SkM) is used to generate artifact-free MRI sketches.
Then a ControlNet-based \cite{controlnet} model MRControlNet (MRCN) is designed to control precise reconstruction details according to the conditions generated in the SkM.

\subsection{Sketcher Module (SkM)}
\label{ssec:subhead}
To ensure reliable conditional information for ControlNet-based pipeline functions, we introduce the sketcher module (SkM) $\phi_1$ to generate an appropriate conditional image $c$ from $M$ and an undersampled image $x_u=F^{-1}(k_u)$, where $F^{-1}$ represents the inverse Fourier transform operator.
\begin{equation}
c=\phi_1(x_u,M).
\label{equ:Skm1}
\end{equation}
In $\phi_1$, a SwinIR\cite{swinir}-based repair model $R$ is first utilized, which can effectively erase aliases in $x_u$ and conduct the LDM focus more on generating realistic details \cite{diffbir}. 
Following the repair module, the data consistency (DC) operation is employed to improve the data fidelity of the condition $c$. 
The DC operation can be expressed as:
\begin{equation}
\begin{aligned}
    c &= DC[R(x_u), k_u, M] \\
      &= F^{-1}\{F[R(x_u)] \cdot (I - M) + k_u \cdot M\},
\end{aligned}
\label{equ:Skm2}
\end{equation}
where $I$ denotes the all-ones matrix, $F$ denotes the forward Fourier transform operator.

\subsection{MRControlNet (MRCN)}
\label{ssec:subhead}

The MRControlNet (MRCN) $\phi_2$ is intended to reconstruct the full-sampled image $\hat{x}$ under the control of the condition image $c$, as illustrated in Fig. \ref{fig:pipeline}. 
Three key parts are involved in MRCN, including 1) MR-VAE, 2) Generator, and 3) Dual-Stage Sampler (DSS).
\begin{equation}
\hat{x}=\phi_2(k_u,M,c).
\label{equ:Skm2}
\end{equation}

\subsubsection{MR-VAE}
\label{sssec:subsubhead}
VAE is an important module to map images into the latent space.
Although many natural image restoration methods \cite{stablesr,diffbir} tend to keep pre-trained VAE weights in stable diffusion (SD)\cite{sd}, such VAEs have not been exposed to MRI images during training and therefore cannot be directly generalized to the MRI domain for accurate MRI detail restoration.
To address this issue, we propose MR-VAE, a VAE that can be used for various MR-related tasks.
MR-VAE is fine-tuned on SD pre-train\cite{sd} to leverage its rich generative prior. 
The encoder and decoder are utilized in the reconstruction process and optimized with three terms of loss functions.
The pixel loss and VGG loss are utilized to reduce the pixel-wise difference; the KL-divergence loss is also introduced to enforce statistical consistency between the learned latent variables and a predefined prior distribution (e.g., Gaussian). 
Additionally, a GAN loss is incorporated to enhance the visual realism of the reconstructed output, improving the learning of sharper edges and more detailed textures\cite{esrgan}.
The total loss can be expressed as:
\begin{multline}
L_{vae}=\mu\cdot||\hat{x}_{vae}-x||_1+\nu\cdot||VGG(\hat{x}_{vae})-VGG({x})||_1\\
+\omega\cdot\mathrm{KL}(N(u, \sigma^2) \parallel N(0, 1)) + \lambda \cdot L_{GAN},
\label{equ:vae}
\end{multline}
where $\mu$, $\nu$, $\omega$ and $\lambda$ represent the weights assigned to each individual loss term.
$\hat{x}_{vae}$ denote the reconstruction result of MR-VAE, $VGG$ denotes the VGG network\cite{vgg}, $u$ and $\sigma$ denote the mean and variance of the reconstruction distribution, respectively, and $L_{GAN}$ represents the adversarial loss introduced by the GAN framework\cite{vqgan}.
As shown in Fig. \ref{fig:vae}, MR-VAE produces more realistic reconstructions on MRI data than the VAE in SD (SD-VAE).

\subsubsection{Generator}
\label{sssec:subsubhead}
 In the generator, a trainable control and a frozen denoiser are utilized.
 The trainable copy of the pre-trained U-Net down-sampler and middle block from SD\cite{sd} are conducted, which preserves generative diffusion prior that trained on a large-scale natural image dataset. 
The condition latent ${c}_{latent}=\mathcal{E}(c)$ and noisy latent $z_t$ are concatenated together as the input of the frozen denoiser. $z_t$ at time step t can be denoted as:
\begin{equation}
z_t=\sqrt{{\alpha}_t}z+\sqrt{1-{\alpha}_t}\epsilon,
\label{equ:zt}
\end{equation}
where $z$ is a latent code encoded from image $x$ ($z=\mathcal{E}(x)$), $\alpha_t$ is from a decreasing sequence where $\alpha_{1:T}\in(0,1]^T$, and the noise $\epsilon\sim \mathcal{N}\left(0,\mathbf{I}\right)$. 
The channel number is increased by the concatenate operation. thus We follow the setup of IRControlNet \cite{diffbir}, appending some extra parameters and then initializing them to zero to avoid noisy gradients in early training steps.
The training loss can be denoted as:

\begin{equation}
L_{MRCN}=\mathbb{E}_{z_t,p,t,\epsilon,c_{latent}}\left[\left|\left|\epsilon-\epsilon_\theta\left(z_t,p,t,c_{latent}\right)\right|\right|_2^2\right],
\label{equ:LMRCN}
\end{equation}
where $p$ is the condition input (i.e., text prompt) and $\epsilon_\theta$ is the learned noise-predicting network. 

\subsubsection{Dual-Stage Sampler (DSS)}
\label{sssec:subsubhead} 
To achieve high-fidelity latent sampling while avoiding the introduction of additional artifacts, a  variant of the spaced DDPM sampler\cite{iddpm} called the Dual-Stage Sampler (DSS) is proposed, inspired by \cite{diffbir} and \cite{u2mrpd}.
Similar to the approach in \cite{u2mrpd}, DSS is designed in two stages based on the time step $t$.
At each time step $t$, the clean latent $\hat{z_0}$ is predicted through \eqref{equ:z0}:
\begin{equation}
\epsilon_t=\epsilon_\theta(z_t,p,t,c_{latent}), \hat{z_0}=\sqrt{\alpha_{t-1}}(\frac{z_t - \sqrt{1 - \alpha_t}\epsilon_t}{\sqrt{\alpha_t}})
\label{equ:z0}
\end{equation}
The division of stages is determined by the constant time $p$, which is set to 200.
Unlike \cite{u2mrpd}, when $t > p\,$, data consistency (DC) is directly applied without random phase to avoid additional uncertainty.
The DC operation \eqref{equ:DC} substitutes the corresponding k-space area in the clean image prediction $\hat{x_0}=\mathcal{D}(\hat{z_0})$ with estimated data $k_u$, which improves data fidelity.
However, it may introduce unwanted artifacts when processing the magnitude images \cite{u2mrpd}.
Thus, DC is only employed in the early sampling steps ($t > p$) where the noise level is higher.
\begin{equation}
\hat{z_0}'=\mathcal{E}[DC(\hat{x_0},k_u,M)]
\label{equ:DC}
\end{equation}
When $t \leq p$, k-space inference guidance that modified from restoration guidance\cite{diffbir} is employed to generate artifact-free and more realistic images. 
In this stage, different from \cite{diffbir}, which optimizes in the pixel domain, the L2 loss between the masked k-space prediction $F(\hat{x_0})$ and the estimated k-space $k_u$, denoted as $\delta$, is calculated (shown in \eqref{equ:CG1}), to adapt to MRI data. 
Then, the generation is guided by $\delta$ through \eqref{equ:CG2}.
\begin{equation}
\delta=\mathcal{E}\{||[F(\hat{x_0})-k_u]\cdot M||_2^2\},
\label{equ:CG1}
\end{equation}
\begin{equation}
\hat{z_0}'=\hat{z_0}-g \nabla_{\hat{z_0}} \delta\,, 
\label{equ:CG2}
\end{equation}
where $g$ is the guidance scale and is set into 0.1. $z_{t-1}$ is sampled from $z_t$ via the standard deviation of Gaussian noise at time t ($\sigma_t$):
\begin{equation}
z_{t-1}=\hat{z_0}'+\sqrt{1-\alpha_{t-1}-\sigma_{t}^2}\cdot\epsilon_t, 
\label{equ:zt-1}
\end{equation}

\vspace{+3mm}
\section{EXPERIMENTAL RESULTS}

\begin{table}[t]
\caption{Quantitative evaluation on the fastmri dataset~\cite{fastmri}}
\vspace{-3mm}
\label{tab:SOTA}
\begin{center}
\fontsize{7.5pt}{10pt}\selectfont 
\setlength{\tabcolsep}{2.5pt} 
\begin{tabular}{ccccccc}
\hline
\hline
\textbf{Method} & \multicolumn{1}{c}{\textbf{PSNR} $\uparrow$} & \multicolumn{1}{c}{\textbf{SSIM} $\uparrow$} & \multicolumn{1}{c}{\textbf{NMSE} $\downarrow$} & \multicolumn{1}{c}{\textbf{LPIPS} $\downarrow$} & \multicolumn{1}{c}{\textbf{rFID} $\downarrow$} & \multicolumn{1}{c}{\textbf{KID} $\downarrow$} \\
\hline
Zero-Filled & 23.9467 & 0.6817 & 0.1052 & 0.4083 & 265.0842 & 0.2543 \\
\hline
U-Net~\cite{fastmri} & 26.7036 & 0.6212 & 0.2845 & 0.3077 & 52.6383 & 0.0282 \\
\hline
E2E-VarNet~\cite{varnet} & \textcolor{blue}{30.0438} & 0.6659 & 0.0300 & 0.2232 & 34.5586 & 0.0165 \\
\hline
SwinMR~\cite{swinmr} & 30.0008 & \textcolor{red}{0.8407} & \textcolor{blue}{0.0259} & 0.2262
& 60.6761 & 0.0372 \\
\hline
Score-MRI~\cite{scoremri} & 29.6848 & 0.7978 & 0.0342 & \textcolor{red}{0.1494}
 & \textcolor{red}{15.7723} & \textcolor{red}{0.0025} \\
\hline
LDPM (Ours) & \textcolor{red}{30.2048} & \textcolor{blue}{0.8042} & \textcolor{red}{0.0247} & \textcolor{blue}{0.1725} & \textcolor{blue}{29.1384} & \textcolor{blue}{0.0116} \\
\hline
\hline
\multicolumn{7}{l}{\fontsize{6.5pt}{7pt}\selectfont \textcolor{red}{Red} and \textcolor{blue}{Blue} indicate the best and second best score.}
\vspace{-4mm}
\end{tabular}
\end{center}
\end{table}

\subsection{Dataset and Implementation Details}
\subsubsection{Dataset}
All the experiments were conducted on the fastMRI brain dataset\cite{fastmri}, including multi-coil brain scans of FLAIR-, T1-, and T2-weighted MR images.  
For each subject, the last $3$ slices were discarded due to poor image quality.
In training, images from 1793 subjects were utilized (a total of $23275$ slices, with $2500$ slices used for validation).
For testing, 72 subjects were involved (a total of $934$ slices). 
All multi-coil MR images were processed into coil-combined magnitude images using the root-sum-of-squares (RSS)\cite{fastmri}.
Cartesian 1D random undersampling mask is utilized in the undersampling reconstruction.

\begin{figure*}
    \centering
    \includegraphics[width=1\linewidth]{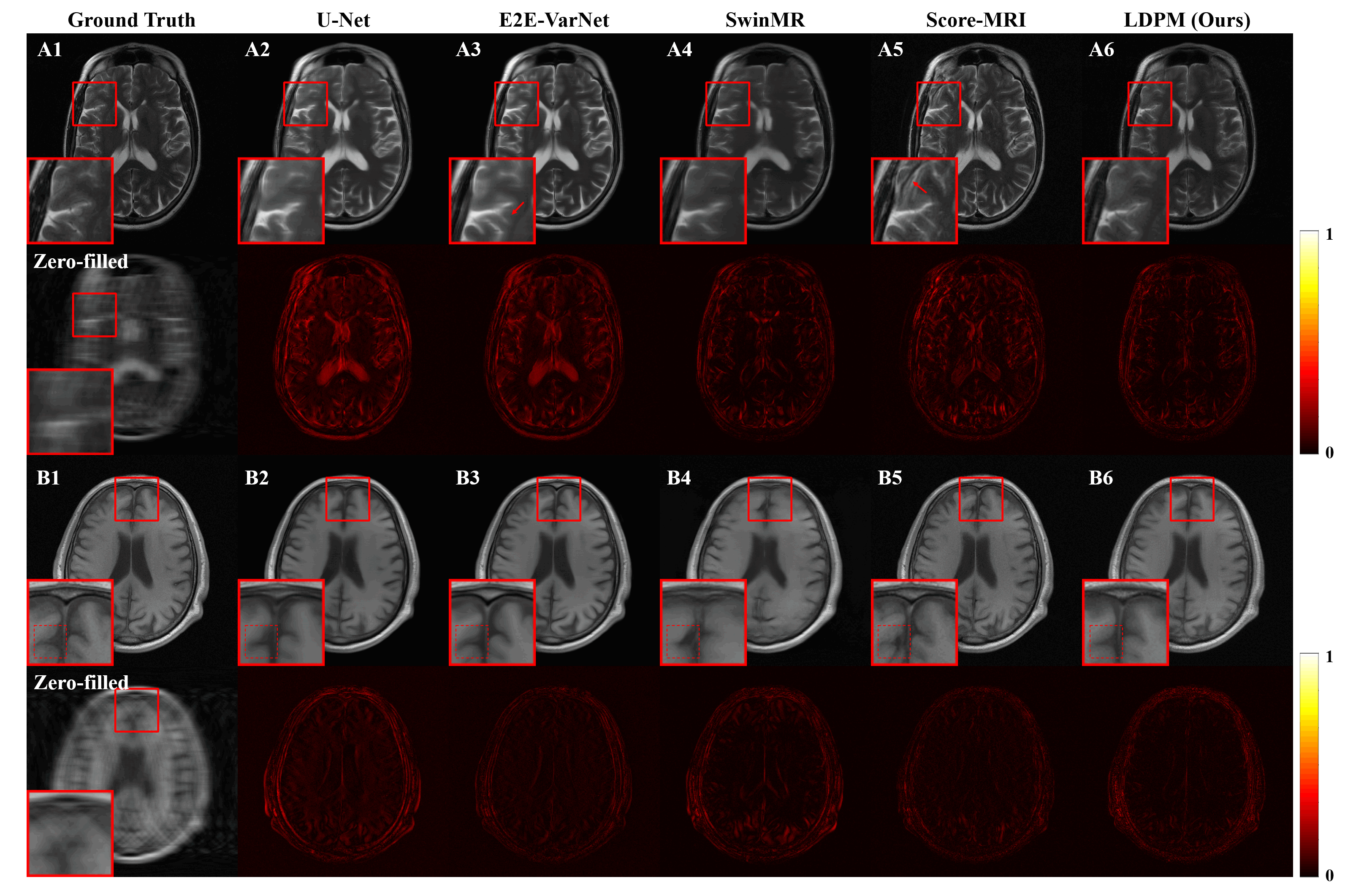}
    \caption{Two sets of examples (A and B) of MRI reconstruction results with 8× acceleration. First and third rows: MRI fully sampled image (GT) and reconstructed results with different undersampled MRI reconstruction methods. Second and fourth rows: zero-filled undersampled images and error maps of each method.
    }
    \label{fig:mainexp}
\end{figure*}

\subsubsection{Implementation Details}
First, the sketcher module (SkM) was trained with a learning rate ($lr$) of $1e-4$ and a batch size ($bs$) of 8.
Then, the MR-VAE was fine-tuned based on the pre-trained stable diffusion 2-1-base\cite{sd} with $lr=1e-5$, $bs=8$, $\mu=1$, $\nu=0.2$, $\omega=1e-6$, and $\lambda=0.65$.
Next, MRControlNet (MRCN) was trained based on MR-VAE with the patch size of $320\times320$, $lr=10e-4$, $bs=16$, and $T=1000$, where MR-VAE was fixed during training.
The downsampling ratio of the MRCN was set to 8, and the text prompt was set to empty.
SkM, MR-VAE, and the generator were trained for 30, 50, and 50 epochs, respectively.


\subsection{Quantitative and Visual Evaluation}
\label{ssec:subhead}
The proposed LDPM method is intuitively and quantitatively evaluated and compared with several classic and state-of-the-art methods, including zero-filled reconstruction, U-Net\cite{fastmri}, E2E-VarNet\cite{varnet}, SwinMR (nPI version)\cite{swinmr}, and score-MRI\cite{scoremri}. 
The official pre-trained models of U-Net\cite{fastmri} and E2E-VarNet\cite{varnet} were utilized.
Both SwinMR\cite{swinmr} and Score-MRI\cite{scoremri} were trained according to their official instructions, with $N=2000$ for Score-MRI, following the official settings.
The quantitative evaluation results, with an acceleration factor (AF) of 8, are shown in Table \ref{tab:SOTA}.
The proposed method outperforms all other algorithms across all metrics. 
While the superior PSNR, SSIM\cite{ssim}, and NMSE demonstrate  high reconstruction fidelity, the lower LPIPS\cite{lpips}, reconstruction FID (rFID)\cite{FID}, and KID\cite{kid} indicate  excellent visual quality. 

\begin{figure*}
    \centering
    \includegraphics[width=1\linewidth]{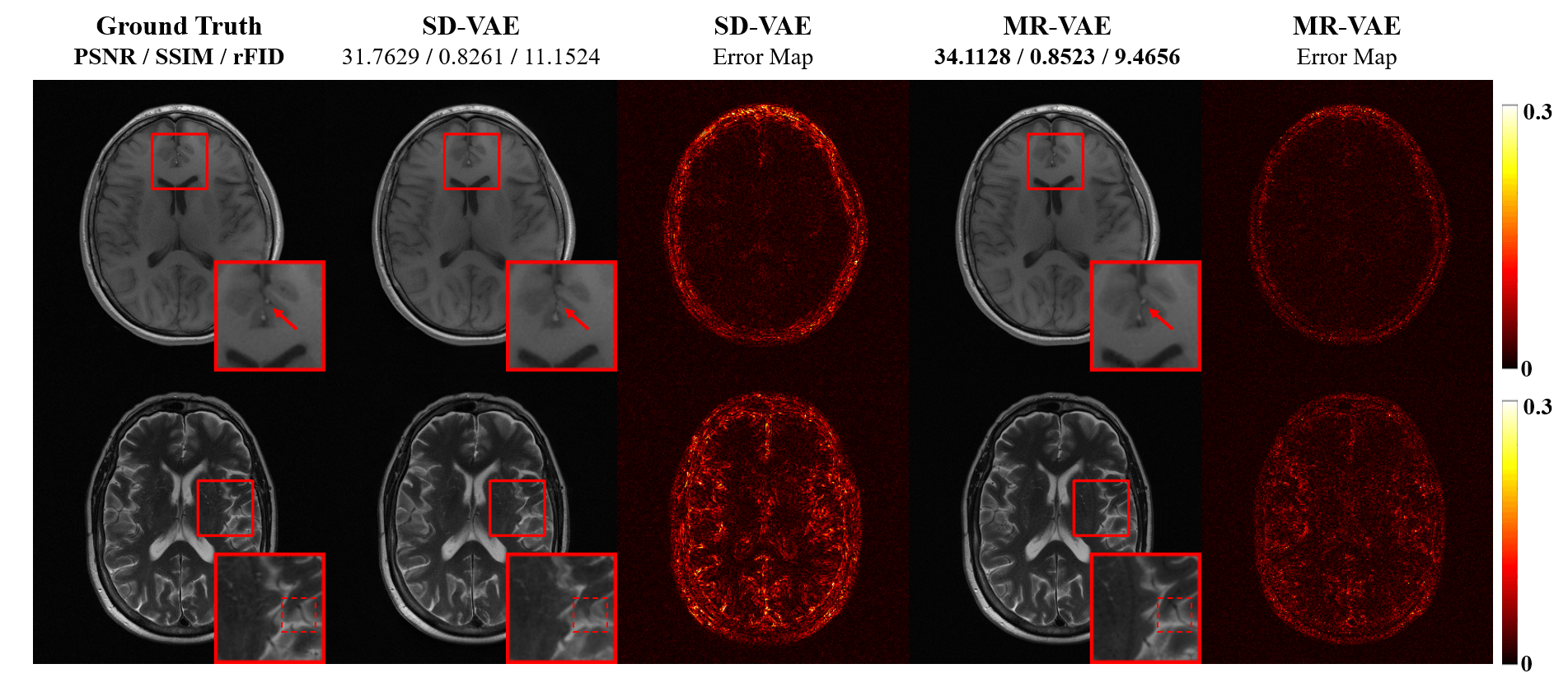}
    \caption{Visual and quantitative comparison of MR-VAE reconstruction results with that of SD-VAE\cite{sd}. Objective metrics, including PSNR, SSIM\cite{ssim}, and rFID\cite{FID}, are computed over the entire testset. First column: ground truths. Second and fourth column: VAE reconstructions and zoomed-in patches. Third / fifth column: corresponding error maps of SD-VAE / MR-VAE reconstructions.}
    \label{fig:vae}
\end{figure*}

\begin{table}[htbp]
\caption{Quantitative evaluation of ablation experiments}
\label{tab:ablation}
\vspace{-2mm}
\begin{center}
\begin{tabular}{ccc}
\hline
\hline
\textbf{Models}&\multicolumn{1}{c}{\textbf{PSNR}}&\multicolumn{1}{c}{\textbf{SSIM}}\\
\hline
LDPM (w/o MR-VAE)&26.2837&0.7166\\
\hline
LDPM (w/o SkM)& 28.2066 & 0.7821\\
\hline
LDPM (w/o DSS)&28.9130&0.7662\\
\hline
LDPM (Ours)&\textbf{30.2048}&\textbf{0.8042}\\
\hline
\hline
\end{tabular}
\label{tab2}
\end{center}
\end{table}

Two sets of test examples of state-of-the-art methods are shown in Fig. \ref{fig:mainexp}.
The reconstruction results obtained from U-Net\cite{fastmri} exhibit a loss of fine details and residual artifacts, as observed in \textbf{A2}, along with occasional brightness distortion, as shown in \textbf{B2}.
E2E-VarNet\cite{varnet} displays smoothing and residual artifacts. 
Specifically, the reconstructed image \textbf{A3} contains additional aliasing textures not present in the ground-truth image \textbf{A1}.
As demonstrated in both examples, SwinMR\cite{swinmr} suffers from significant detail loss, leading to reconstructions that lack texture.
Score-MRI\cite{scoremri} tends to generate images with unrealistic details in a stochastic manner, as seen in both examples, which could potentially interfere with medical diagnoses.
The proposed LDPM method achieves superior, artifact-free reconstruction results with realistic details, showcasing the robust generalization capabilities of the latent fusion prior.

\subsection{Effectiveness of Key Components} 
\label{ssec:subhead}
The effectiveness of each component in the proposed method, including MR-VAE, the transformer-based sketcher module (SkM), and Dual-Stage Sampler (DSS), was examined through a series of ablation studies.
Results are shown in Table \ref{tab:ablation}.
It can be seen that the removal of each component will lead to losses in PSNR and SSIM\cite{ssim}, proving that each component is effective.

In particular, MR-VAE is compared with SD-VAE\cite{sd}, a widely used pre-trained VAE on natural images, as shown in Fig. \ref{fig:vae}.
Quantitative evaluation with PSNR, SSIM\cite{ssim} and rFID\cite{FID} was conducted on the $934$ testing images. 
MR-VAE demonstrates superior performance, as evidenced by higher SSIM, PSNR, and lower rFID scores compared to SD-VAE\cite{sd}, highlighting its enhanced suitability for MRI reconstruction tasks.
A closer look at the zoomed-in detail patches, indicated by arrows and dashed boxes, reveals that MR-VAE generates more realistic details, whereas SD-VAE \cite{sd} often overlooks fine structures, which may lead to misinterpretation and degrade the quality of reconstructed images.

\begin{table}[t]
\caption{PSNR evaluation across modalities, field strengths, and on out of distribution content}
\label{tab:psnr}
\centering
\fontsize{6pt}{9pt}\selectfont 
\setlength{\tabcolsep}{1.25pt} 
\begin{tabular}{c!{\color{gray!50}\vrule}ccccc!{\color{gray!50}\vrule}cc!{\color{gray!50}\vrule}c}
\hline
\hline
\multirow{2}{*}{Method} & \multicolumn{5}{c!{\color{gray!50}\vrule}}{Modality} 
& \multicolumn{2}{c!{\color{gray!50}\vrule}}{Field Strength} & Content \\
 & \textbf{FLAIR} & \textbf{T1} & \textbf{T1PRE} & \textbf{T1POST} & \textbf{T2} & \textbf{1.5 Tesla} & \textbf{3 Tesla} & \textbf{fastMRI knee} \\
\hline
\hline
\textbf{Zero-filled} & 24.1651 & 24.5031 & 24.8114 & 23.9130 & 22.8707 & 23.6512 & 24.2579 & 18.4156 \\
\hline
\textbf{U-Net\cite{fastmri}} & 25.4063 & 25.2059 & 27.7707 & 27.3613 & 26.1399 & 22.7030 & 22.7880 & 13.6225 \\
\hline
\textbf{E2E-Varnet\cite{varnet}} & \textcolor{blue}{28.3468} & 29.4912 & 29.9914 & \textcolor{blue}{30.6448} & \textcolor{red}{29.2099} & \textcolor{red}{30.2732} & 29.8024 & 19.8002 \\
\hline
\textbf{SwinMR\cite{swinmr}} & 28.3113 & \textcolor{blue}{30.4111} & \textcolor{blue}{30.8969} & 30.6220 & 28.1087 & \textcolor{blue}{29.8921} & \textcolor{blue}{30.1153} & 18.2742 \\
\hline
\textbf{Score-MRI\cite{scoremri}} & 27.2958 & 29.7731 & 30.2489 & 30.6061 & 27.7811 & 29.3245 & 30.0641 & \textcolor{blue}{22.242} \\
\hline
\textbf{Ours (LDPM)} & \textcolor{red}{28.4670} & \textcolor{red}{30.5707} & \textcolor{red}{31.1840} & \textcolor{red}{30.8185} & \textcolor{blue}{28.5416} & \textcolor{blue}{30.1087} & \textcolor{red}{30.3060} & \textcolor{red}{23.3653} \\
\hline
\hline
\multicolumn{9}{l}{\fontsize{7pt}{8pt}\selectfont \textcolor{red}{Red} and \textcolor{blue}{Blue} indicate the best and second-best scores, respectively.}
\vspace{-4mm}
\end{tabular}
\end{table}

\subsection{Robustness Evaluation}
\label{ssec:subhead}

To evaluate the robustness of the proposed method, experiments on model adaptability are conducted across modalities, field strengths, and an out-of-distribution dataset. 
For cross-modality evaluation, 100 images were randomly selected from the testset for each modality. 
To assess the model's adaptability to data from different field strengths, the testset was divided into images measured under 1.5 Tesla and 3 Tesla field strength environments, with 479 and 455 images, respectively. Additionally, to evaluate the generalization performance, 189 slices from 7 individuals were randomly selected from the fastMRI knee dataset\cite{fastmri} with AF of 8 to evaluate the performance of various methods on unseen data.

As presented in Table \ref{tab:psnr}, LDPM consistently outperforms the existing state-of-the-art methods across these diverse settings, demonstrating its robustness against measurement shifts, and reliable generalization capability in handling unseen scenarios, highlighting the outstanding adaptability and effectiveness of LDPM.

\section{Conclusion}
In this work, a latent diffusion prior based MRI reconstruction method LDPM is proposed.
An MRI prior-enhanced MR-VAE and a latent space-adapted Dual-Stage Sampler are proposed to reduce the latent space transformation loss and improve the reconstruction fidelity, respectively. 
The LDPM method outperforms state-of-the-art methods and demonstrates the  potential of applying latent diffusion priors in future MRI reconstruction studies.
\vspace{+
10mm}

\begin{spacing}{0.98}
\bibliographystyle{IEEEtran}
\bibliography{abrvbib}
\end{spacing}

\end{document}